\documentclass[letterpaper]{article} 
\usepackage{aaai25}  
\usepackage{times}  
\usepackage{helvet}  
\usepackage{courier}  
\usepackage{xcolor}
\usepackage[hyphens]{url}  
\usepackage{graphicx} 
\usepackage{natbib}  
\usepackage{caption} 
\usepackage{algorithm}
\usepackage{algorithmic}
\usepackage{booktabs}
\usepackage{fontawesome}

\usepackage{amssymb}

\usepackage{newfloat}
\usepackage{listings}
\DeclareCaptionStyle{ruled}{labelfont=normalfont,labelsep=colon,strut=off} 
\floatstyle{ruled}
\newfloat{listing}{tb}{lst}{}
\floatname{listing}{Listing}
\title{Column-Oriented Datalog on the GPU}
\author {
    Yihao Sun\textsuperscript{\rm 1}, 
    Sidharth Kumar\textsuperscript{\rm 2}, 
    Thomas Gilray\textsuperscript{\rm 3}, 
    Kristopher Micinski\textsuperscript{\rm 1}
}
\affiliations {
    \textsuperscript{\rm 1}Syracuse University\\
    \textsuperscript{\rm 2}University of Illinois at Chicago\\
    \textsuperscript{\rm 3}Washington State University\\
    \{ysun67, kkmicins\}@syr.edu, sidharth@uic.edu, thomas.gilray@wsu.edu
}

\usepackage{bibentry}

\newcommand{\tool}{\textsc{FVlog}}
\newcommand{\GDlog}{\textsc{GDlog}}

\begin{document}

\maketitle

\begin{abstract}
Datalog is a logic programming language widely used in knowledge representation and reasoning (KRR), program analysis, and social media mining due to its expressiveness and high performance. 
Traditionally, Datalog engines use either row-oriented or column-oriented storage. Engines like VLog and Nemo favor column-oriented storage for efficiency on limited-resource machines, while row-oriented engines like Souffl\'e use advanced data structures with locking to perform better on multi-core CPUs.
The advent of modern datacenter GPUs, such as the NVIDIA H100 with its ability to run over 16k threads simultaneously and high memory bandwidth, has reopened the debate on which storage layout is more effective. This paper presents the first column-oriented Datalog engines tailored to the strengths of modern GPUs. We present \tool{}, a CUDA-based Datalog runtime library with a column-oriented GPU datastructure that supports all necessary relational algebra operations. Our results demonstrate over $200\times$ performance gains over SOTA CPU-based column-oriented Datalog engines and a $2.5\times$ speedup over GPU Datalog engines in various workloads, including KRR.
\end{abstract}

%

\section{Introduction}
\label{sec:introduction}
Datalog~\cite{abiteboul1995foundations} has become a de-facto standard for reasoning across a breadth of fields, including deductive databases~\cite{saenz2011deductive}, program analysis~\cite{bravenboer2009doop}, business analytics \cite{aref2015extending}, and knowledge representation~\cite{nenov2015rdfox}. For example, popular applications of RDF, such as OWL 2 RL ontologies~\cite{motik2019maintenance} with SWRL rules~\cite{horrocks2004swrl}, are easily transliterated into Datalog rules. A Datalog engine then executes these rules to a fixed point, materializing a database for subsequent querying. 
While early systems such as OWLIM~\cite{kolovski2010optimizing} and WebPIE~\cite{urbani2010owl} were challenged by slow performance, modern Datalog engines enable scaling useful queries to internet-scale datasets~\cite{ajileye2022materialisation}.

High-performance Datalog engines such as RDFox~\cite{motik2014parallel} and Souffl\'e \cite{jordan2016souffle} are typically designed for CPU-based hardware, storing tuples in row-based format with minimally-locking data structures such as B-Trees and tries \cite{jordan2019brie,jordan2019specialized}. 
However, row-based storage can face scalability challenges on modern CPUs, especially those with multiple Core Chiplet Dies (CCDs) that do not share caches. Accessing entire rows can lead to suboptimal cache alignment and increased memory access latency, particularly for tasks such as joins and indexing, which frequently access only a subset of columns. Additionally, these systems are not entirely lock-free, which can further limit scalability as core counts increase. Our benchmarks show that both engines' performance saturates at 32 cores and declines rapidly when scaling up to 64 cores.

Some CPU-based in-memory Datalog reasoners, such as VLog and Nemo~\cite{urbani2016vlog, nemo2023}, adopt a column-oriented tuple representation~\cite{abadi2008column} along with compression techniques such as run-length encoding~\cite{robinson1967rle}. On CPU-based systems, the column-oriented approach trades space for time, enabling these engines to run on machines with limited RAM.

However, the performance claim between these two storage layouts shifts with the emergence of new hardware, particularly datacenter GPUs. These SIMD-like superchips, such as the NVIDIA H100, can execute more than 16 k threads simultaneously and are supported by 96 GB of high-bandwidth memory (HBM)~\cite{hbm}, and have significant potential to accelerate data-intensive workloads.
Such massively parallel systems demand higher memory locality and fully lock-free code, making it challenging to migrate existing row-oriented Datalog engines to this hardware.
In contrast, the column-oriented storage layout, with its smaller tuple size, naturally fits the SIMD architecture. This layout has already proven to be more cache-friendly and offers better memory locality for SIMD-enhanced CPU-based systems, as demonstrated in the database community~\cite{ailamaki1999dbmss, zukowski2008dsm}, and is widely used in high-performance OLAP databases such as DuckDB~\cite{raasveldt2019duckdb} and MonetDB/X100~\cite{boncz2005monetdb}.

Due to the data-intensive nature of Datalog and the similarities between GPUs and SIMD processors, we believe that column-oriented storage could be a better fit for modern datacenter GPUs. In this paper, we present \tool{}, a column-oriented Datalog Engine backend designed for modern GPUs. Our contributions are as follows.

\begin{itemize}
\item We present the first-ever column-oriented relation-backing representation for GPU-based Datalog engines.
\item We implement \tool{}, a CUDA-based GPU Datalog runtime library that supports efficient join, copy, deduplication, fixpoint computation, and other primitive operations for modern high-performance Datalog engines.
\item  We perform a thorough evaluation. Our results show over $200\times$ speedup compared to CPU-based column-oriented systems and $2.5\times$ faster performance than other GPU prototypes in both standard Datalog and knowledge graph reasoning workloads.   
\end{itemize}


\section{Preliminaries}
\label{sec:preliminaries}

\paragraph{Datalog and RDF}
Datalog restricts Prolog's to positive Horn clauses.
It consists of a set of clauses in the form $H \leftarrow~ B_1, ... , B_n$, where the head $H$ is derived if all body clauses $B_{1..n}$ are satisfied. A Datalog engine infers new facts (materializing an Intensional Database, IDB) from the rules and ground facts (also called the Extensional Database, EDB) provided as input.
For example, transitive closure can be represented in following rules: 
\[
\begin{array}{lcl}
    \textit{Reach}(x,y)  & \!\! \leftarrow & \!\!  \textit{Edge}(x, y).   \\
    \textit{Reach}(x, z) & \!\! \leftarrow & \!\! \textit{Edge}(x, y), \textit{Reach}(y, z). \\
\end{array}
\]

Datalog is chain-forward, recursively evaluating all rules to infer all possible facts until a fixed point is reached.
One compelling application of Datalog is in materializing RDF knowledge graphs.
RDF is a powerful framework for representing information about resources in the form of a directed, labeled graph. Edges in a knowledge graph may be represented via RDF triples using a predicate that relates two objects; for example, a knowledge graph for a family relationship may be represented as:
\[
\begin{array}{lcr}
    <\texttt{Alice} & \textit{:parentOf} & \texttt{Bob}> \\
    <\texttt{Larry} & \textit{:parentOf} & \texttt{Alice}> 
\end{array}
\]
One straightforward way to solve RDF reasoning problem in Datalog is by encoding the RDF triples as binary relation facts, then run recursive Datalog queries on it to materialize all information can be derived from the RDF graph.
For example, above RDF triples can be encoded in Datalog as:
\[
\begin{array}{lccr}
    \textit{parentOf} ~( & \!\!\!\!\! \texttt{Alice} &\!\!,\!\!& \texttt{Bob}) \\
    \textit{parentOf} ~( & \!\!\!\!\! \texttt{Larry} &\!\!,\!\!& \texttt{Alice})
\end{array}
\]
Following transitive closure like Datalog rules can be used to materialize  the ancestor relationship in the family graph:
\[
\begin{array}{lcl}
    \textit{ancestor}(x, y)  & \!\! \leftarrow & \!\!  \textit{parentOf}(x, y).   \\
    \textit{ancestor}(x, z) & \!\! \leftarrow & \!\! \textit{parentOf}(x, y), \textit{ancestor}(y, z). \\
\end{array}
\]
After above program reach the fixed point, the \textit{ancestor} relation will contain all ancestor relationship in the family graph, we can directly query the \textit{ancestor} to get the result.

\begin{figure}
    \centering
    \includegraphics[width=\linewidth]{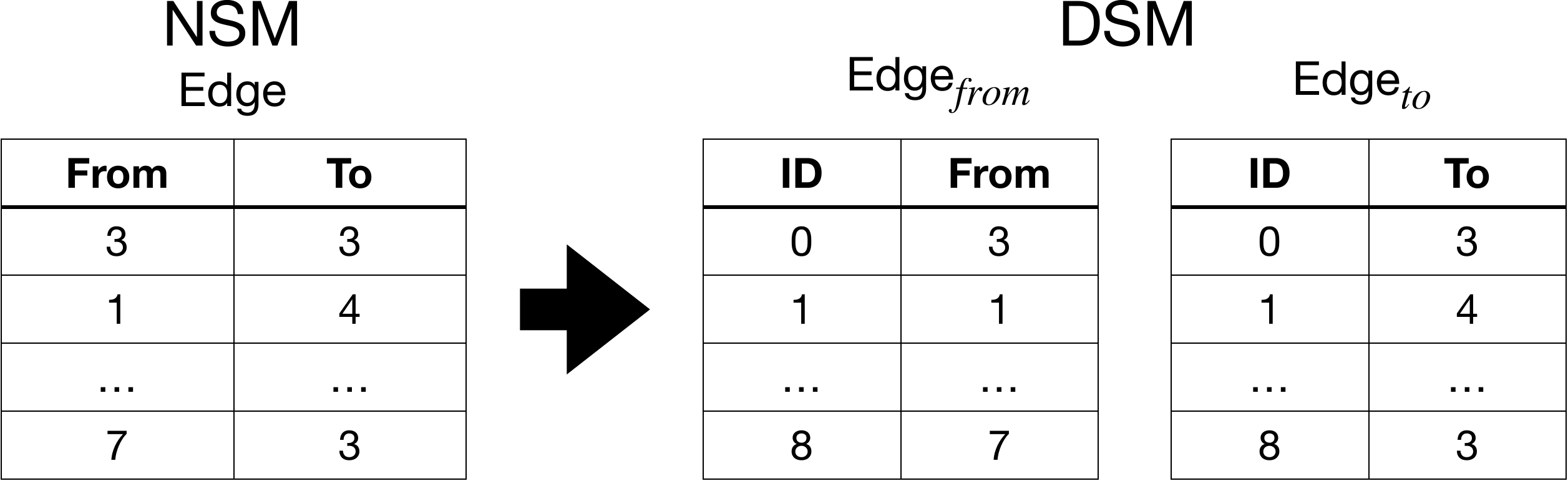}
    \caption{Converting \textit{Edge} relation from NSM to DSM.}
    \label{fig:dsm}
\end{figure}

\paragraph{Decomposed Storage Model (DSM)}
Database records are traditionally stored as rows of n-ary tuples in a horizontal layout known as the N-ary Storage Model (NSM). Even today, most database management systems (DBMS) utilize NSM. However, some research demonstrated that storing database records via vertical columns could offer better performance~\cite{weyl1975modular}. This approach inspired the Decomposed Storage Model (DSM).
Figure~\ref{fig:dsm} shows an example of converting a NSM relation to DSM. An \textit{Edge} relation is broken into two binary relations: $\textit{Edge}_{\textit{from}}$ and $\textit{Edge}_{\textit{to}}$. Each of these binary relations includes an additional \textit{ID} column that stores the original row number. For instance, the second row of the original relation, $(1,4)$, is decomposed into $(1,1)$ in $\textit{Edge}_{\textit{from}}$ and $(1,4)$ in $\textit{Edge}_{to}$. This \textit{ID} column, also known as the \textit{surrogate column}, is essential for reconstructing the entire relation during join operations.
Formally, a n-ary relation $R(x_0, ..., x_n)$ is decomposed to:
\[
R_0(id, x_0), R_1(id, x_1), ..., R_n(id, x_n)
\]
In the DSM model, accessing all values in the same column becomes straightforward—one simply needs to access each decomposed column relation. If the entire row is required, a join operation can be performed to combine the columns using the surrogate column. An n-ary DSM relation's row can be reconstructed as follows:
\[
\Pi_{\neq id}(\textit{R}_0 \bowtie_{id} \textit{R}_1 \bowtie_{id} ... \bowtie_{id} \textit{R}_n)
\]
Transposing data into DSM facilitates vectorization, allowing multiple tuples to be processed in parallel, and improves cache efficiency with smaller tuple sizes \cite{ailamaki2001weaving}. These benefits are similar to those achieved by converting an Array of Structures (AoS) into a Structure of Arrays (SoA) in GPGPU programming~\cite{pennycook2013investigation}. SoA is considered the best practice for GPUs due to improved memory coalescing and cache performance~\cite{colease}.
Given these advantages, recent surveys suggest that DSM could also enhance performance in GPU-based databases~\cite{zeng2023empirical}, applying the same principles that benefit GPU workloads to DBMS.

While the DSM model offers excellent read performance, it often increases memory overhead due to the additional \textit{id} column in each binary relation. Modern DSM-based databases, such as MonetDB/X100~\cite{boncz2005monetdb} and C-Store~\cite{stonebraker2018c}, mitigate this overhead by storing tuples in lexicographic order and employing column compression techniques. However, the separation of each row's storage and the associated datastructure overhead make the write operation more expensive. This overhead is particularly problematic in Datalog, where the materialized IDB is often orders-of-magnitude larger than the input EDB.

\section{Column-Oriented Relations on the GPU}
\label{sec:datastructure}
\begin{figure}
    \centering
    \includegraphics[width=\linewidth]{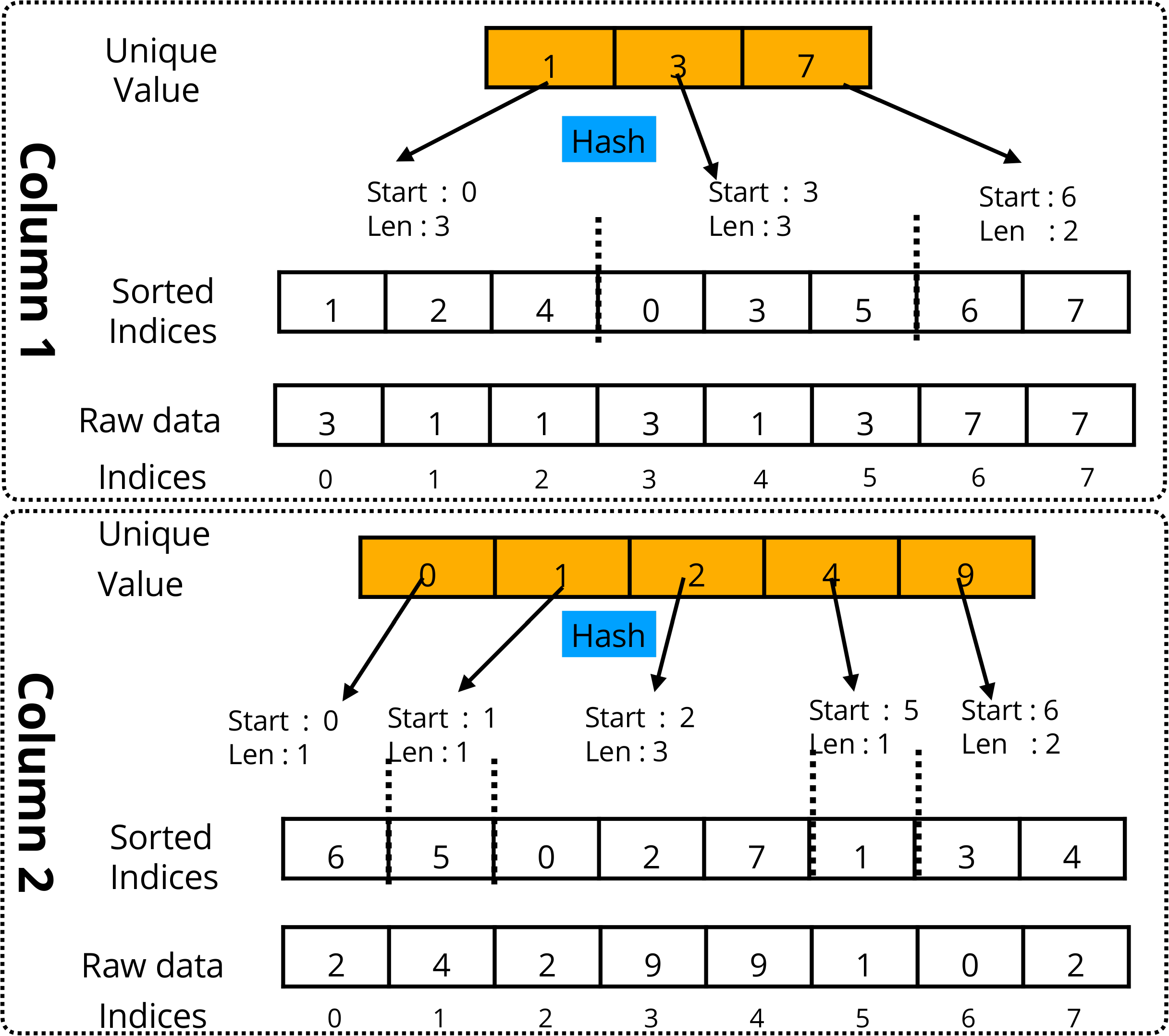}
    \caption{ \textit{Reach} stored in column-oriented layout on GPU.}
    \label{fig:datastructure}
\end{figure}

In our approach, the relations are stored using the DSM, where each relation is decomposed into columns, and each column shares an identical datastructure. Figure~\ref{fig:datastructure} illustrates our approach, particularized to the case of 2-ary relation. The column data structure consists of three main components: a \emph{raw data array}, \emph{sorted indices}, and a \emph{unique hashmap}. The bottom of each column in Figure~\ref{fig:datastructure} shows the raw data array, which stores the actual data of the relation in 32-bit integers ordered by logical tuple insertion time.
A benefit of decomposing raw tuple rows into separate columns is that doing so facilitates aligned access on modern GPUs; GPU cores are typically 32-bit computation units organized into warps. To maximize parallel performance, all GPU threads in the same warp must execute the same instruction and access memory in a coalesced manner, with each thread accessing consecutive memory addresses \cite{cuda}.
When data is organized in a row-oriented manner, any relation with more than one attribute would exceed the size of a 32-bit integer, causing each GPU core to access non-consecutive memory addresses when processing data in the same column. These misaligned accesses significantly degrade throughput. By contrast, using column-oriented storage ensures that each column is stored in consecutive memory addresses, facilitating aligned access and aiding throughput. 


The middle section of each column in Figure~\ref{fig:datastructure} represents the \textit{sorted indices}, aligned with the surrogate column in the DSM model. Each element in this array is an offset into the raw data array, ordered by the values in that data. For instance, in column $2$ of Figure~\ref{fig:datastructure}, the $0^{th}$ value in the sorted indices is $6$, pointing to the $6^{th}$ element in the raw data array, which is 0—the smallest value in this column. This sorted index acts as a database index, allowing quick location of corresponding value ranges during join operations, thus avoiding a full scan of the raw data array.

Pure binary search-based join operations on sorted indices can be inefficient on GPUs due to in-warp thread divergence caused by heavy conditional branching. To address this, we use a hybrid indexing approach that combines sorted indices with a \textit{unique hashmap} at the top of each column in Figure~\ref{fig:datastructure}. This design further enhances the efficiency of join operations. This hashmap stores run-length encoded values from the columns, with each unique value as a key. The hashmap value is a pair comprising the starting offset in the sorted indices and the count of occurrences. For example, in column 1 of Figure~\ref{fig:datastructure}, the value 1 appears three times in the raw data array, starting at the $0^{th}$ element in the sorted indices, so the hashmap stores the pair (0, 3) for the key 1. This index structure enables fast hash joins and increases throughput.

We avoid using a pure hashmap for indexing because, unlike CPU, GPU hashmaps typically use linear probing and open addressing to resolve hash collisions. This approach, as seen in popular implementations such as cuCollection \cite{cucollection}, optimizes cache performance by ensuring continuous memory allocation. In Datalog, columns often contain many repeated values, leading to frequent hash collisions and degraded performance. Our benchmarks against GPUJoin \cite{shovon2023towards} demonstrate that our duplicated eliminated hybrid indexing approach is more efficient for real-world Datalog workloads.

Although our approach is inspired by the CPU-based column-oriented reasoner VLog, our design makes several important departures to optimize performance on modern GPUs, which we now describe.



\paragraph*{Continuous Memory Layout}
To avoid re-evaluating known facts during fixpoint computation, Datalog implementations often adopt the semi-na\i"ve evaluation. This method optimizes computation by managing each relation in three versions: \textit{full} (all tuples), \textit{delta} (tuples from the most recent iteration), and \textit{new} (tuples generated in the current iteration), focusing on delta facts to streamline the process. However, this algorithm requires a write-intensive merge operation between \textit{delta} and \textit{full} in each iteration.

Due to the costly write operations associated with DSM relations, Datalog systems such as VLog avoid merging the \textit{delta} generated in each iteration into the \textit{full} relation. Instead, each datastructure object contains only the tuples generated within the same iteration. All relational algebra operations on the \textit{full} relation are broken down into a series of operations on the \textit{delta} in each iteration. While this design saves time during insertion, it introduces sequential overhead.
When the number of iterations is low, this overhead is manageable on CPU-based systems. However, in real-world Datalog recursive queries, such as transitive closure on large graphs, hundreds of iterations are common. Excessive iterations cause the fragmented \textit{full} relation to be scattered across memory, leading to poor memory locality and cache performance, which significantly increases overhead during joins.

The mitigating solution used by VLog involves \textit{on-demand concatenation} to construct index during join operations. However, this approach remains unsuitable for GPUs. First, while on-demand concatenation helps reduce overhead during joins, it does not address the need to deduplicate \textit{new} relation data with the fragmented \textit{full} relation during semi-na\"ive evaluation. Second, this approach introduces memory management overhead due to the frequent, large memory allocations and deallocations required for temporary consolidated data structures in each iteration.

To address these challenges and better align with GPU architecture, we take the opposite approach. Instead of avoiding the overhead of data insertion, we accept it to maintain continuity in the raw data. This continuity ensures better data locality and cache performance, enabling more effective parallelism across the entire Datalog evaluation process.
Moreover, due to the high memory bandwidth of GPUs, the parallel insertion of delta tuples into the \textit{full} relation is not as expensive as on CPUs. Time-consuming operations such as sorting, scanning, and hash table construction can be efficiently parallelized \cite{satish2010fast, green2012gpu, green2021hashgraph}. Therefore, in \tool{}, we eagerly merge all delta tuples into the \textit{full} relation to maintain the memory continuity and ensure better data locality.

\paragraph*{Uncompressed Raw Data}
To save memory bandwidth, CPU-based column-oriented database systems use RLE compression on the raw data array. These systems are typically optimized for CPU in-memory processing. In contrast, we target datacenter GPUs, which offer large VRAM capacities of up to 192 GB, making raw column compression less beneficial. Thus, we compress only the indices and not the raw data. This approach avoids the overhead of decompressing raw data during parallel relational algebra operations such as joins and copies---wherein thousands of GPU threads need to iterate over raw tuples simultaneously---and maximizing parallelism and memory bandwidth of the GPU.

\paragraph{Schedule multiple rules per iteration}
VLog employs a one-rule-per-step variant of semi-na\"ive evaluation, where in each iteration, only one rule is applied. In this setup, if two Datalog rules contribute to the same relation in the same iteration--for example, in transitive closure, where both rules could contribute to the \textit{Reach}--it results in two separate merge operations. In contrast, the original semi-na\"ive evaluation would have both rules contribute to the same relation simultaneously, requiring only a single merge.
This is not a proper design for GPUs. In \tool{}, we adopt the original semi-na\"ive evaluation design.

\section{Relational Algebra Operators}

There are multiple ways to implement Datalog queries, such as using Binary Decision Diagrams (BDD) \cite{whaley2004cloning}, SMT solvers \cite{hoder2011muz}, and Answer Set Programming (ASP) \cite{calimeri2017dlv}. However, state-of-the-art high-performance Datalog engines such as Souffl\'e~\cite{aref2015design} achieve their high throughput via relational algebra kernels~\cite{ceri1986translation}.
In this approach, a Datalog query is translated into a series of extended positive relational algebra ($\mathcal{R\!A}^{+}$) \cite{ullman1983principles} operations, including join ($\bowtie$), 
selection ($\sigma$), 
projection ($\Pi$), 
and set union ($\cup$), 
along with an extended closure operator ($\mathcal{O}$), which computes the fixpoint.
For example, to compute the transitive closure query, we can use the following $\mathcal{R\!A}^{+}$ expression:
\[
\begin{array}{lcl}
\textit{Reach} & = & \mathcal{O}(\Pi_{x, z}(\textit{Edge} \bowtie_{y} \textit{Reach})  ~ \cup ~ \textit{Reach} )
\end{array}
\]

We now discuss the implementation of these $\mathcal{R\!A}^{+}$ operations in a column-oriented fashion optimized for GPUs. 

\paragraph*{Projection and Selection}
In \tool{}, the projection operator processes only surrogate columns instead of entire rows. The selection operator operates directly on the raw data array of the targeted column. Both RA operators can be implemented efficiently using NVIDIA’s CCCL library.

\begin{algorithm}[t]
    \caption{Binary Join on hash indexed DSM relations}
    \begin{algorithmic}[1]
        \STATE \textbf{Input:} $R_A, R_B$ are two DSM relations columns
        \STATE \textbf{Output:} the result of join stored in $R_C$ 
        \STATE ranges, matched$_A$ $\leftarrow$ allocate(size of $R_A$)
        \FOR{$c$ in $R_A$, $x$ is id of $c$ \textbf{parallel}}
            \IF {$R_{B}$.hashmap[$c$]}
            \STATE ranges[x] $\leftarrow$~ $R_{B}$.hashmap[$c$]
            \STATE matched$_A$[x] $\leftarrow$ x
            \ELSE
            \STATE ranges[x] $\leftarrow$~ ($\emptyset$, $\emptyset$)
            \STATE matched$_A$[x] $\leftarrow \emptyset$
            \ENDIF
        \ENDFOR
        \STATE filter out empty position in ranges and matched$_A$
        \STATE total\_size $\gets$ \textbf{parallel} sum sizes of all ranges
        \STATE $R_C \leftarrow$ \textbf{parallel} allocate(total\_size)
        \STATE pos\_buf $\gets$ exclusive\_scan(size of each ranges)
        \FOR{$n$ from 0 to R\_C.size - 1 \textbf{parallel}}
            \STATE upper\_bound $\gets$ \\ ~~~~ j + 1 $\le$ pos\_buf.size ? pos\_buf[$j+1$] : total\_size
            \STATE find $j$ such that pos\_buf[$j$] $\leq$ $n$ $<$ pos\_buf[$j+1$]
            \STATE $A_{id} \gets$ matched$_A$[$j$]
            \STATE $B_{id} \gets$ $R_B$.sorted\_id[ranges[$j$].start + \\ ~~~~~~~~~~~~($n$ - pos\_buf[$j$])]
            \STATE write ($A_{id}$, $B_{id}$) to $R_C$[$n$]
        \ENDFOR
    \end{algorithmic}
    \label{alg:binaryjoin}
\end{algorithm}

\paragraph*{Join}
Unlike the typical join operator in traditional row-oriented databases, we do not materialize the full join result within the join operator itself. Instead, we only return the matched surrogate columns of the join candidate relations. Join results are materialized only during a projection operation when the result columns are actually needed. This approach helps avoid unnecessary memory allocation for the full join result, saving both memory and computation time.

\begin{figure}
    \centering
    \includegraphics[width=\linewidth]{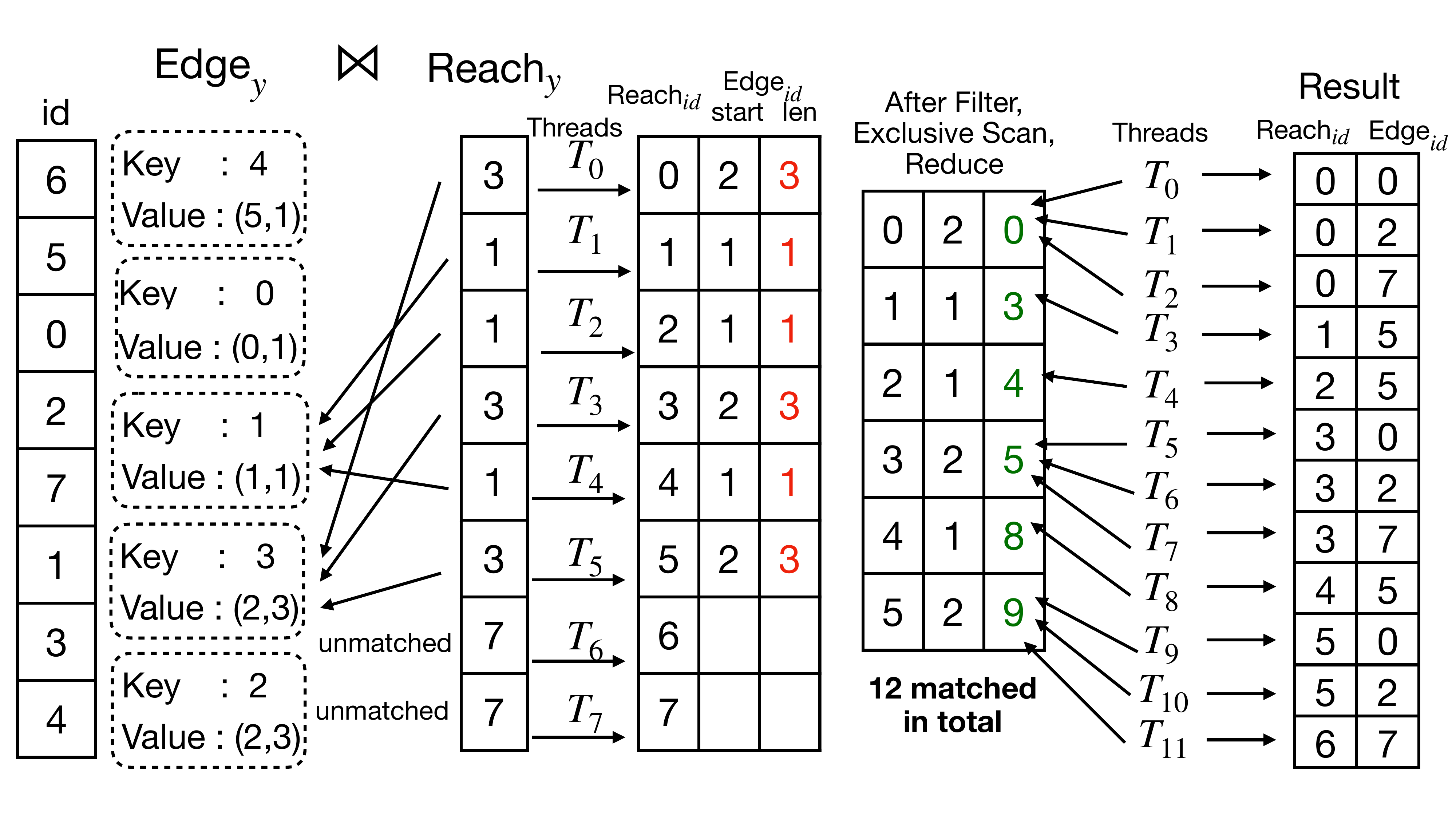}
    \caption{Example of $\textit{Reach} \bowtie_{y} \textit{Edge}$ paralleled on GPU.}
    \label{fig:binaryjoin}
\end{figure}

Algorithm~\ref{alg:binaryjoin} shows the implementation of join. The entire join can be divided into two main phases: computing join size (line 3 to line 16) and writing join results (line 17 to line 23). The first phase of the join process is illustrated in the left half of Figure~\ref{fig:binaryjoin}. In this phase, each GPU thread parallelly iterates over the data column of the decomposed $\textit{Reach}_{y}$ ($R_A$ in the algorithm). For each value, the thread queries the hashmap of $\textit{Edge}_{y}$ ($R_B$ in the algorithm) to find the matched surrogate values and the corresponding tuple ID ranges that share the same raw data value.
Some values in $\textit{Reach}_{y}$ may not have a match in $\textit{Edge}_{y}$—for example, the bottom ``7'' in the Figure~\ref{fig:binaryjoin}. In such cases, these unmatched values are filtered out by a filter function in line 16 of the algorithm.
Next, by applying a parallel \texttt{reduce} function (line 14 of the algorithm) on the lengths of the matched ranges (indicated in red in the Figure~\ref{fig:binaryjoin}), we compute the total number of join results, which in this case is 12. We then allocate memory for the results based on this computed size. Separating the join size computation phase from the result writing phase, rather than performing everything in a single loop (as is common in CPU-based engines), allows us to allocate memory for the join results in advance
, which enables each thread to write to the join result without any lock contention.

There are two ways to collect the join result after computing the matched ranges. The first method involves parallel iteration over all matched ranges, where each thread writes a different number of tuples depending on the length of each range, which can cause data skew. For example, in Figure~\ref{fig:binaryjoin}, the first matched range has a length of 3, while the second has only 1. The second method involves dividing the workload based on the output result, ensuring that each thread writes the same number of tuples, thereby avoiding data skew. However, this method requires extra searches within each thread to find the corresponding matched range, therefore most of CPU-based engines usually prefer the first method. However, most GPU algorithms favor the second style because reducing thread divergence improves performance when the thread count is large. In our implementation, we chose the second method for writing the join result.

We begin the second phase by performing a parallel exclusive prefix sum (line 16) on the lengths of each range identified in the previous phase to generate a result offset buffer, where each element represents the starting position in the result. These computed positions are marked in green in Figure~\ref{fig:binaryjoin}. For example, nothing precedes the first range, so it has a result offset of ``0''. The third range has two preceding ranges with sizes of 3 and 1, so the resulting offset for the third range is 4. Then, we parallel iterate over all positions in the proclaimed result memory. Within each thread, a sequential search (which can be accelerated by binary search) identifies the corresponding matched position in the ranges, and the join result is written to the output buffer. For instance, thread 2 processes position number 2 in the result; by performing a binary search on the green buffer computed earlier, it determines that position 2 is between 0 and 3, matching the first range, which corresponds to the second to fifth values in the ID column of $\textit{Edge}_{y}$. According to lines 19 and 20 in the algorithm, we compute that \textit{Reach} ID 0 and \textit{Edge} ID 2 should be written to the result buffer.



\paragraph*{Union and Deduplication}
Implementing set union becomes more complex in DSM because the deduplication process during a join typically requires simultaneous access to entire rows.
To effectively handle deduplication, we further extend $\mathcal{RA}^{+}$ with the difference ($-$) operator, which removes tuples from the left relation that have a matching tuple on the right. This extension allows us to efficiently manage deduplication.
For example, the join operation in the transitive closure can be translated as:
\[
\begin{array}{lcl}
\textit{New} & = & \textit{Edge}_1 \bowtie_{y} \textit{Reach}_0 \\
\Delta & = & \textit{New} - (\textit{New} \bowtie_{x} \textit{Reach}_0 \bowtie_{id, y} \textit{Reach}_1) \\
\textit{Reach}_{0,1} & = & \mathcal{O} \left(\Pi_{x} (\Delta) \cup \textit{Reach}_0, \Pi_{y} (\Delta) \cup \textit{Reach}_1\right)
\end{array}
\]
Note that although we use the symbol $\cup$ here since the tuples have already been deduplicated, the set union operation is effectively a simple concatenation.

The join pattern revealed in this approach aligns with the classic \textit{triangle join} problem, where three relations are joined to form a cycle. The complexity of evaluating such joins is tied to the AGM (Algebraic Graph Model) bound~\cite{atserias2013size}, which provides a worst-case estimate of the size of join results based on the sizes of the input relations. The AGM bound suggests that when data skew is present, any conventional binary join plan can become prohibitively expensive--a significant challenge for Datalog engines that rely on such plans.
To illustrate this, consider the join within a deduplication process, which can be expressed in the Horn Clause as follows:
\[
\textit{Dedup}(x, y) \leftarrow ~ \textit{New}(x, y), \textit{Reach}_0(id, x), \textit{Reach}_1(id, y)
\]
Here, \textit{New} represents the new tuples generated from the join between \textit{Reach} and \textit{Edge}. $x,y,\textit{id}$ form a join circle.

A classic solution is to use a trie-based join algorithm, such as Leapfrog Triejoin (LFTJ) \cite{veldhuizen2014leapfrog}, widely used in CPU-based Datalog engines such as LogicBlox. However, LFTJ is unsuitable for GPUs because its Leapfrog search step is inherently sequential. Some other solutions such as generic join \cite{ngo2014skew} and its column-oriented variant free join \cite{wang2023free} have been proposed to address the triangle join problem in the context of traditional CPU-based databases. However, the recursive nature of free join makes it particularly challenging to implement on GPU.

\begin{algorithm}
    \caption{Deduplication in \tool{} for a 2-arity relation}
    \begin{algorithmic}[1]
        \STATE \textbf{Input:} $\textit{New}$(x, y), S(id, x), T(id, y) are input relations
        \STATE \textbf{Output:} Q is bitmap for matched $\textit{New}$
        \FOR{$a$ in $\textit{New}_{x}$ \textbf{parallel}}
            \IF{$S$.hashmap[$a$]}
                \STATE $range_{x}$ $\leftarrow$ $S$.hashmap[$a$]  
            \ELSE 
            \STATE $range_{x}$ $\leftarrow$ ($\emptyset$,$\emptyset$)
            \ENDIF
        \ENDFOR
        \STATE match$_{id}$ = range($R$.size)
        \STATE do the same for $\textit{New}_{y}$ and $T$ generate range$_{y}$
        \STATE remove $i$ from match$_{id}$ if either range is empty
        \FOR{$i$ in match$_{id}$ \textbf{parallel}}
            \STATE if range$_{x}$[$i$] and range$_{y}$[$i$] are overlapped, Q[$i$]=\TRUE
        \ENDFOR
    \end{algorithmic}
    \label{alg:join}
\end{algorithm}

Therefore, we take a different approach to handling triangle joins during deduplication. Instead of using tries, we tailor the process to our GPU-friendly datastructure. The details are presented in Algorithm~\ref{alg:join}. For simplicity, we demonstrate the deduplication process for a 2-arity relation, though the same approach can be extended to n-arity relations.

Algorithm~\ref{alg:join} begins by computing the normal hash join between \textit{New} and all the other columns in the full relation (the $S$ and $T$ in the algorithm).
Before joining surrogate columns (lines 10-12), tuples with empty matched surrogate column ranges are marked for early elimination, reducing unnecessary computation. To prevent thread divergence and warp serialization, especially in the presence of data skew, the two join operations are separated into distinct parallel loops. After marking tuples (lines 13-15), another parallel loop efficiently checks for matches in the surrogate column.

While this approach may be less efficient in terms of memory usage compared to a solution such as LFTJ, since it requires additional buffers equivalent in size to the newly generated relation, managing these buffers can be challenging and often requires techniques such as best-fit allocation \cite{shore1975external}. However, this lock-free design is particularly well-suited for GPUs, and the massive parallel speedup it offers makes the additional memory cost worthwhile.


\section{Evaluation}
We evaluated the performance of \tool{} through three key comparisons. 
First, we compare \tool{} with CPU-based column-oriented Datalog engines, VLog and Nemo, to demonstrate how GPU-optimized data structures can accelerate column-oriented Datalog. Next, we compare \tool{} with GPU-based row-oriented Datalog prototypes, GPUJoin and \GDlog{}, to highlight the superior performance of column-oriented storage on GPUs. Finally, we validate \tool{} by running the LUBM scenario in ChaseBench to demonstrate its applicability in knowledge graph reasoning.

\subsection{Experimental Environment}
All our experiments were conducted on a server equipped with an AMD EPYC 9534 and an NVIDIA H100. The AMD EPYC 9534 features 64 cores and 128 threads, supported by 500 GB of memory with a memory bandwidth of 0.43 TB/s. The NVIDIA H100 GPU includes 16,896 CUDA cores and 80 GB of HBM3 memory, offering up to 3.3 TB/s of memory bandwidth. The server runs Ubuntu 22.04 and GCC 11. For VLog, we used Rulewerk, a Java wrapper for VLog that provides additional language features. For Nemo, we utilized version 0.5.1. We used Souffl'e version 2.4.1, with maximum multithreading and compiler optimizations. The RDFox version used was 7.1a, with all CPU threads enabled. All GPU tools were compiled with NVC++ in NVHPC 24.1.

\begin{table}[htbp]
    \centering
    \resizebox{\linewidth}{!}{%
    \begin{tabular}{lcccccc}
        \toprule
        Dataset & Size & \tool{} & VLog & Nemo & Souffl\'e & RDFox\\
        \midrule
        vsp\_finan & 552,020 & 7.52   &  4403  & 2172     & 151.5 & 257  \\
        fc\_ocean & 409,593  & 0.31   &  169.7  & 151.9    & 13.13 & 19.2 \\
        SF.cedge  & 223,001 & 1.80   &  1121   & 298.9    & 56.52 & 117 \\
        fe\_body  & 163,734 & 1.85   &  173.9  & 555.4    & 48.18 & 126 \\
        CA-HepTH  & 51,971 & 0.55   &  313.7  & 147.7    & 20.12 & 20.1 \\ 
        fe\_sphere & 49,152 & 0.92   &  582.7  & 160.8     & 48.12 & 63.8 \\
        \bottomrule
    \end{tabular}}
    \caption{Running time (Second) of Same Generation Query. \tool{} and \GDlog{} are executed on NVIDIA H100, Nemo and souffl\'e are exectuetd on AMD EPYC 9534 (Genoa). }
    \label{tab:sg}
\end{table}

\subsection{Column-Oriented Datalog Comparison}
We first compared the performance of \tool{} with two CPU-based column-oriented Datalog engines, VLog and Nemo, using a simple yet representative Same Generation (SG) Datalog query. This query is a common pattern in Datalog reasoning and demands substantial computation time. 
\[
\begin{array}{lcl}
    \textit{SG}(x,y) & \leftarrow & \textit{Edge}(p,x), ~ \textit{Edge}(p, y), ~ x \neq y. \\
    \textit{SG}(x, y) & \leftarrow & \textit{Edge}(a, x), ~ \textit{SG}(a,b), ~ \textit{Edge}(b,y), x \neq y.
\end{array}
\]
Results are presented in Table~\ref{tab:sg}. The first column lists the names of the graphs used in this experiment, all of which come from the SparseSuite \cite{davis2011sparsesuite} dataset. The size of each input data is reported in the second column. These graphs are real-world graphs extracted from diverse areas such as road systems, simulations, and SAT solving. This diversity ensures that our benchmark of the Datalog engines is unbiased. Columns three through five lists the running times for the benchmark candidates, while the last two columns include the state-of-the-art industrial Datalog engines, Souffl\'e and RDFox, as a reference.

The results demonstrate the significant performance gains achieved by running Datalog on GPUs. Under similar storage layouts (all candidates are column-oriented), \tool{} outperforms VLog and Nemo by a large margin. In all test cases, \tool{} on the H100 is at least more than 150 times faster than VLog and Nemo on AMD Genoa. Notably, on the \textit{vsp\_finan} dataset, which is a comparatively large input graph containing 552,020 edges, \tool{} is 584 times faster than VLog and 288 times faster than Nemo. Even when compared to Souffl\'e and RDFox, which are optimized multicore Datalog engines, \tool{} on the datacenter GPU still shows a significant performance advantage. On the \textit{vsp\_finan} dataset, \tool{} is 20 times faster than Souffl\'e, demonstrating the superior performance in large datasets.

\subsection{Comparing \tool{} to SOTA Datalog Engines}

\begin{table}[htbp]
    \centering
    \resizebox{\linewidth}{!}{%
    \begin{tabular}{lcccccc}
        \toprule
        Dataset     & \tool{} & \GDlog{} & GPUJoin & Souffl\'e & RDFox \\
        \midrule
        vsp\_finan  & 7.94   & 21.91  & 63.89  & 239.3 & 269 \\
        fe\_ocean   & 10.07  & 23.36  & \faBug{}      & 292.2 & 507 \\
        usroads     & 9.55   & 17.53  & 57.89  & 243.1 &  268 \\
        com-dblp    & 3.35   & 14.30  &  \faBug{}     & 233.0 & 569 \\
        Gnutella31  & 1.2    & 3.76   & 7.82   & 96.82 & 373 \\
        fe\_sphere  & 0.53   & 0.93   & 1.16   & 25.02 & 25.1 \\
        \bottomrule
    \end{tabular}}
    \caption{Running time (Second) of Transitive Closure Query on \tool{} and \GDlog{} are executed on NVIDIA H100, souffl\'e are exectuetd on AMD EPYC 9534.  \faBug{}  indicates a non-OOM crash observed running GPUJoin.}
    \label{tab:tc}
\end{table}

To validate whether the column-oriented storage layout outperforms the row-oriented layout on GPUs, we compare \tool{} with two GPU-based row-oriented Datalog prototypes, GPUJoin~\cite{shovon2023towards} and \GDlog{}~\cite{sun2023gdlog}. We use the transitive closure query mentioned in Section~\ref{sec:preliminaries} as the benchmark and also include Souffl\'e and RDFox as a reference. The running time results are shown in Table~\ref{tab:tc}. Due to some bugs in the code, we were unable to obtain results for GPUJoin on the \textit{fe\_ocean} and \textit{com-dblp} datasets.
The results indicate that all GPU-based engines show significant performance improvements over the CPU-based engines. In sum, \tool{} is on average $2.5\times$ faster than \GDlog{} and $5.7\times$ faster than GPUJoin.

Our investigation showed that the comparatively low performance of GPUJoin is due to the use of a hash map for indexing used in this engine. Additionally, GPUJoin's need to compress entire rows into single 32-bit integers, restricting it to 2-arity relations, limits its versatility compared to \tool{}.
\tool{}'s advantage over \GDlog{}, despite both using hybrid indexing, lies in its column-oriented storage, which improves data locality and memory bandwidth utilization. The per-column processing in the DSM model further simplifies expensive operations like tuple sorting, enabling the use of parallel radix sort instead of merge sort, which is more efficient on GPUs.


\begin{table}[htbp]
    \centering
    \resizebox{\linewidth}{!}{%
    \begin{tabular}{lccccc}
        \toprule
        Dataset & \tool{}(C) & \tool{}(G) & Nemo & VLog & RDFox \\
        \midrule
        010  & 0.15   & 0.01   & 0.65    & 1.46  & 0.44 \\
        100 & 0.71   & 0.03   & 6.34    & 5.02   & 4.98 \\
        01K  & 6.47   & 0.16   & 62.32   & 165.8 & 56.8 \\
        \bottomrule
    \end{tabular}}
    \caption{TGD reasoning time (seconds) comparison of different Datalog engine on LUBM.
             \tool{}(C) runs on AMD EPYC 9534 CPU, \tool{}(G) runs on H100 GPU.}
    \label{tab:lubm}
\end{table}

\subsubsection{Benchmark Knowledge Representation and Reasoning}
After benchmarking the basic Datalog reasoning queries, we further evaluated the performance of \tool{} on a KRR workload. We selected tuple-generating dependency (TGD) queries (excluding existential rules) on the LUBM dataset from the ChaseBench ~\cite{benedikt2017benchmarking}, a widely used dataset for evaluating Datalog-based KRR systems. The queries used in this test include both ST-TGD and T-TGD and were sourced from the example repository of Nemo \cite{nemo-example}. The results are presented in Table~\ref{tab:lubm}. 
The first column lists the names of the sub-datasets used in this test, with the size of the input data increasing from top to bottom. The third column shows the running time of \tool{} on a H100 GPU, while the fourth and fifth columns display the running times of Nemo and VLog on an 64 cores EPYC 9543.
In the last column, we also put the RDFox industrial row-oriented reasoner as a reference.
By comparing the running times, we can conclude that combining the power of GPUs with performance-aware data structures enables \tool{} to significantly improve Datalog materialization times for KRR workloads. The maximum speedup achieved is close to $300\times$ in the largest input dataset compared to traditional CPU-based systems, Nemo and RDFox. This test, being more copy-intensive than previous join-heavy benchmarks, reveals that RDFox, even with 64 CPU cores, does not significantly outperform sequential reasoners. We attribute this to the memory-bound nature of the queries and the lack of data locality in row-oriented storage, which limits parallel performance.

To further investigate the contributions of performance-oriented data structures versus the benefits derived from the incredible memory bandwidth and core size of GPUs, we also developed a CPU variant of \tool{}. This variant employs similar data structures but leverages Intel's latest oneTBB \cite{onetbb} to utilize the multicore resources on a datacenter CPU instead of GPU threading. The running times of the CPU version of \tool{} are listed in the second column of Table~\ref{tab:lubm}.
The results show that the GPU version of \tool{} is at least $15\times$ faster than the CPU version. Considering that the memory bandwidth of the H100 is nearly $7.9\times$ times larger than the EPYC 9534, this indicates that the workload is memory-bound and that the performance gains on the GPU are primarily due to its high memory bandwidth. Additionally, when comparing the running times of the CPU version of \tool{} with Nemo and VLog, we observe around a $9.6\times$ speedup on the largest dataset. 
This suggests that performance-oriented data structures also contribute to about half of the overall improvement.

\section{Conclusion and Future Work}

In this paper, we demonstrate that a column-oriented storage layout is superior to a row-oriented layout on GPUs for Datalog processing.
However, we also admitted that the uncompressed design of \tool{} and the persistence of surrogate columns result in higher memory usage. Despite this, the trend towards hardware with larger memory capacities and higher memory bandwidths makes \tool{} well-suited for future advancements.
Looking ahead, another promising direction is to develop a cluster version of \tool{} to utilize the fast interconnects and advanced load balancing available in modern HPC environments. This would help overcome current memory size limitations and further enhance scalability and performance in distributed settings.

\section{Acknowledgements}
This work was funded in part by NSF PPoSS large grants CCF-2316159 and CCF-2316157. 
This material is based upon work supported by the Defense Advanced Research Projects Agency (DARPA) under Contract No. N66001-21-C-4023. Any opinions, findings and conclusions or recommendations expressed in this material are those of the author(s) and do not necessarily reflect the views of DARPA.

\bibliography{aaai25}

\begin{thebibliography}{51}
\providecommand{\natexlab}[1]{#1}

\bibitem[{Abadi, Madden, and Hachem(2008)}]{abadi2008column}
Abadi, D.~J.; Madden, S.~R.; and Hachem, N. 2008.
\newblock Column-stores vs. row-stores: how different are they really?
\newblock In \emph{Proceedings of the 2008 ACM SIGMOD international conference on Management of data}, 967--980.

\bibitem[{Abiteboul, Hull, and Vianu(1995)}]{abiteboul1995foundations}
Abiteboul, S.; Hull, R.; and Vianu, V. 1995.
\newblock \emph{Foundations of databases}, volume~8.
\newblock Addison-Wesley Reading.

\bibitem[{Ailamaki et~al.(2001)Ailamaki, DeWitt, Hill, and Skounakis}]{ailamaki2001weaving}
Ailamaki, A.; DeWitt, D.~J.; Hill, M.~D.; and Skounakis, M. 2001.
\newblock Weaving Relations for Cache Performance.
\newblock In \emph{VLDB}, volume~1, 169--180.

\bibitem[{Ailamaki et~al.(1999)Ailamaki, DeWitt, Hill, and Wood}]{ailamaki1999dbmss}
Ailamaki, A.; DeWitt, D.~J.; Hill, M.~D.; and Wood, D.~A. 1999.
\newblock DBMSs on a modern processor: Where does time go?
\newblock In \emph{VLDB'99, Proceedings of 25th International Conference on Very Large Data Bases, September 7-10, 1999, Edinburgh, Scotland, UK}, 266--277.

\bibitem[{Ajileye and Motik(2022)}]{ajileye2022materialisation}
Ajileye, T.; and Motik, B. 2022.
\newblock Materialisation and data partitioning algorithms for distributed RDF systems.
\newblock \emph{Journal of Web Semantics}, 73: 100711.

\bibitem[{Aref et~al.(2015{\natexlab{a}})Aref, Kimelfeld, Pasalic, and Vasiloglou}]{aref2015extending}
Aref, M.; Kimelfeld, B.; Pasalic, E.; and Vasiloglou, N. 2015{\natexlab{a}}.
\newblock Extending datalog with analytics in LogicBlox.
\newblock In \emph{Proceedings of the 9th Alberto Mendelzon International Workshop on Foundations of Data Management}.

\bibitem[{Aref et~al.(2015{\natexlab{b}})Aref, Ten~Cate, Green, Kimelfeld, Olteanu, Pasalic, Veldhuizen, and Washburn}]{aref2015design}
Aref, M.; Ten~Cate, B.; Green, T.~J.; Kimelfeld, B.; Olteanu, D.; Pasalic, E.; Veldhuizen, T.~L.; and Washburn, G. 2015{\natexlab{b}}.
\newblock Design and implementation of the LogicBlox system.
\newblock In \emph{Proceedings of the 2015 ACM SIGMOD International Conference on Management of Data}, 1371--1382.

\bibitem[{Atserias, Grohe, and Marx(2013)}]{atserias2013size}
Atserias, A.; Grohe, M.; and Marx, D. 2013.
\newblock Size bounds and query plans for relational joins.
\newblock \emph{SIAM Journal on Computing}, 42(4): 1737--1767.

\bibitem[{Benedikt et~al.(2017)Benedikt, Konstantinidis, Mecca, Motik, Papotti, Santoro, and Tsamoura}]{benedikt2017benchmarking}
Benedikt, M.; Konstantinidis, G.; Mecca, G.; Motik, B.; Papotti, P.; Santoro, D.; and Tsamoura, E. 2017.
\newblock Benchmarking the chase.
\newblock In \emph{Proceedings of the 36th ACM SIGMOD-SIGACT-SIGAI Symposium on Principles of Database Systems}, 37--52.

\bibitem[{Boncz, Zukowski, and Nes(2005)}]{boncz2005monetdb}
Boncz, P.~A.; Zukowski, M.; and Nes, N. 2005.
\newblock MonetDB/X100: Hyper-Pipelining Query Execution.
\newblock In \emph{Cidr}, volume~5, 225--237.

\bibitem[{Bravenboer and Smaragdakis(2009)}]{bravenboer2009doop}
Bravenboer, M.; and Smaragdakis, Y. 2009.
\newblock Strictly declarative specification of sophisticated points-to analyses.
\newblock In \emph{Proceedings of the 24th ACM SIGPLAN conference on Object oriented programming systems languages and applications}, 243--262.

\bibitem[{Calimeri et~al.(2017)Calimeri, Fusc{\`a}, Perri, and Zangari}]{calimeri2017dlv}
Calimeri, F.; Fusc{\`a}, D.; Perri, S.; and Zangari, J. 2017.
\newblock I-DLV: the new intelligent grounder of DLV.
\newblock \emph{Intelligenza Artificiale}, 11(1): 5--20.

\bibitem[{Ceri, Gottlob, and Lavazza(1986)}]{ceri1986translation}
Ceri, S.; Gottlob, G.; and Lavazza, L. 1986.
\newblock Translation and optimization of logic queries: The algebraic approach.
\newblock In \emph{Proceedings of the 12th International Conference on Very Large Data Bases}, 395--402.

\bibitem[{cuCollection(2024)}]{cucollection}
cuCollection. 2024.
\newblock cuCollections (cuco), an open-source, header-only library of GPU-accelerated, concurrent data structures.
\newblock \url{https://github.com/NVIDIA/cuCollections}.
\newblock Accessed: 2024-08-30.

\bibitem[{Davis and Hu(2011)}]{davis2011sparsesuite}
Davis, T.~A.; and Hu, Y. 2011.
\newblock The university of Florida sparse matrix collection.
\newblock \emph{ACM Trans. Math. Softw.}, 38(1).

\bibitem[{Green(2021)}]{green2021hashgraph}
Green, O. 2021.
\newblock HashGraph—Scalable hash tables using a sparse graph data structure.
\newblock \emph{ACM Transactions on Parallel Computing (TOPC)}, 8(2): 1--17.

\bibitem[{Green, McColl, and Bader(2012)}]{green2012gpu}
Green, O.; McColl, R.; and Bader, D.~A. 2012.
\newblock GPU merge path: a GPU merging algorithm.
\newblock In \emph{Proceedings of the 26th ACM international conference on Supercomputing}, 331--340.

\bibitem[{Hoder, Bj{\o}rner, and De~Moura(2011)}]{hoder2011muz}
Hoder, K.; Bj{\o}rner, N.; and De~Moura, L. 2011.
\newblock $\mu$Z--an efficient engine for fixed points with constraints.
\newblock In \emph{Computer Aided Verification: 23rd International Conference, CAV 2011, Snowbird, UT, USA, July 14-20, 2011. Proceedings 23}, 457--462. Springer.

\bibitem[{Horrocks et~al.(2004)Horrocks, Patel-Schneider, Boley, Tabet, Grosof, Dean et~al.}]{horrocks2004swrl}
Horrocks, I.; Patel-Schneider, P.~F.; Boley, H.; Tabet, S.; Grosof, B.; Dean, M.; et~al. 2004.
\newblock SWRL: A semantic web rule language combining OWL and RuleML.
\newblock \emph{W3C Member submission}, 21(79): 1--31.

\bibitem[{{Intel}(2024)}]{onetbb}
{Intel}. 2024.
\newblock {oneAPI Threading Building Blocks (oneTBB)}.
\newblock \url{https://github.com/oneapi-src/oneTBB}.
\newblock Accessed: 2024-08-30.

\bibitem[{Ivliev et~al.(2023)Ivliev, Ellmauthaler, Gerlach, Marx, Mei{\ss}ner, Meusel, and Kr{\"{o}}tzsch}]{nemo2023}
Ivliev, A.; Ellmauthaler, S.; Gerlach, L.; Marx, M.; Mei{\ss}ner, M.; Meusel, S.; and Kr{\"{o}}tzsch, M. 2023.
\newblock Nemo: First Glimpse of a New Rule Engine.
\newblock In Pontelli, E.; Costantini, S.; Dodaro, C.; Gaggl, S.; Calegari, R.; Garcez, A.~D.; Fabiano, F.; Mileo, A.; Russo, A.; and Toni, F., eds., \emph{Proceedings 39th International Conference on Logic Programming (ICLP 2023)}, volume 385 of \emph{EPTCS}, 333--335.

\bibitem[{{JEDEC}(2021)}]{hbm}
{JEDEC}. 2021.
\newblock {High Bandwidth Memory (HBM) DRAM}.
\newblock \url{https://www.jedec.org/document_search?search_api_views_fulltext=jesd235}.
\newblock Accessed: 2024-08-30.

\bibitem[{Jordan, Scholz, and Suboti{\'c}(2016)}]{jordan2016souffle}
Jordan, H.; Scholz, B.; and Suboti{\'c}, P. 2016.
\newblock Souffl{\'e}: On synthesis of program analyzers.
\newblock In \emph{Computer Aided Verification: 28th International Conference, CAV 2016, Toronto, ON, Canada, July 17-23, 2016, Proceedings, Part II 28}, 422--430. Springer.

\bibitem[{Jordan et~al.(2019{\natexlab{a}})Jordan, Suboti{\'c}, Zhao, and Scholz}]{jordan2019brie}
Jordan, H.; Suboti{\'c}, P.; Zhao, D.; and Scholz, B. 2019{\natexlab{a}}.
\newblock Brie: A specialized trie for concurrent datalog.
\newblock In \emph{Proceedings of the 10th International Workshop on Programming Models and Applications for Multicores and Manycores}, 31--40.

\bibitem[{Jordan et~al.(2019{\natexlab{b}})Jordan, Suboti{\'c}, Zhao, and Scholz}]{jordan2019specialized}
Jordan, H.; Suboti{\'c}, P.; Zhao, D.; and Scholz, B. 2019{\natexlab{b}}.
\newblock A specialized B-tree for concurrent datalog evaluation.
\newblock In \emph{Proceedings of the 24th symposium on principles and practice of parallel programming}, 327--339.

\bibitem[{KBS(2024)}]{nemo-example}
KBS. 2024.
\newblock Nemo Examples and Benchmarks.
\newblock \url{https://github.com/knowsys/nemo-examples/blob/main/chasebench/lubm/}.
\newblock Accessed: 2024-08-30.

\bibitem[{Kolovski, Wu, and Eadon(2010)}]{kolovski2010optimizing}
Kolovski, V.; Wu, Z.; and Eadon, G. 2010.
\newblock Optimizing enterprise-scale OWL 2 RL reasoning in a relational database system.
\newblock In \emph{International Semantic Web Conference}, 436--452. Springer.

\bibitem[{Motik et~al.(2019)Motik, Nenov, Piro, and Horrocks}]{motik2019maintenance}
Motik, B.; Nenov, Y.; Piro, R.; and Horrocks, I. 2019.
\newblock Maintenance of datalog materialisations revisited.
\newblock \emph{Artificial Intelligence}, 269: 76--136.

\bibitem[{Motik et~al.(2014)Motik, Nenov, Piro, Horrocks, and Olteanu}]{motik2014parallel}
Motik, B.; Nenov, Y.; Piro, R.; Horrocks, I.; and Olteanu, D. 2014.
\newblock Parallel materialisation of datalog programs in centralised, main-memory RDF systems.
\newblock In \emph{Proceedings of the AAAI Conference on Artificial Intelligence}, volume~28.

\bibitem[{Nenov et~al.(2015)Nenov, Piro, Motik, Horrocks, Wu, and Banerjee}]{nenov2015rdfox}
Nenov, Y.; Piro, R.; Motik, B.; Horrocks, I.; Wu, Z.; and Banerjee, J. 2015.
\newblock RDFox: A highly-scalable RDF store.
\newblock In \emph{The Semantic Web-ISWC 2015: 14th International Semantic Web Conference, Bethlehem, PA, USA, October 11-15, 2015, Proceedings, Part II 14}, 3--20. Springer.

\bibitem[{Ngo, R{\'e}, and Rudra(2014)}]{ngo2014skew}
Ngo, H.~Q.; R{\'e}, C.; and Rudra, A. 2014.
\newblock Skew strikes back: new developments in the theory of join algorithms.
\newblock \emph{Acm Sigmod Record}, 42(4): 5--16.

\bibitem[{NVIDIA(2024{\natexlab{a}})}]{colease}
NVIDIA. 2024{\natexlab{a}}.
\newblock CUDA Best Practice Guide: Coalesced Access to Global Memory.
\newblock \url{https://docs.nvidia.com/cuda/cuda-c-best-practices-guide/index.html#coalesced-access-to-global-memory}.
\newblock Accessed: 2024-08-30.

\bibitem[{NVIDIA(2024{\natexlab{b}})}]{cuda}
NVIDIA. 2024{\natexlab{b}}.
\newblock CUDA Programming Guide: Programming Models.
\newblock \url{https://docs.nvidia.com/cuda/cuda-c-programming-guide/index.html#programming-model}.
\newblock Accessed: 2024-08-30.

\bibitem[{Pennycook et~al.(2013)Pennycook, Hammond, Wright, Herdman, Miller, and Jarvis}]{pennycook2013investigation}
Pennycook, S.~J.; Hammond, S.~D.; Wright, S.~A.; Herdman, J.; Miller, I.; and Jarvis, S.~A. 2013.
\newblock An investigation of the performance portability of OpenCL.
\newblock \emph{Journal of Parallel and Distributed Computing}, 73(11): 1439--1450.

\bibitem[{Raasveldt and M{\"u}hleisen(2019)}]{raasveldt2019duckdb}
Raasveldt, M.; and M{\"u}hleisen, H. 2019.
\newblock Duckdb: an embeddable analytical database.
\newblock In \emph{Proceedings of the 2019 International Conference on Management of Data}, 1981--1984.

\bibitem[{Robinson and Cherry(1967)}]{robinson1967rle}
Robinson, A.~H.; and Cherry, C. 1967.
\newblock Results of a prototype television bandwidth compression scheme.
\newblock \emph{Proceedings of the IEEE}, 55(3): 356--364.

\bibitem[{S{\'a}enz-P{\'e}rez, Caballero, and Garc{\'\i}a-Ruiz(2011)}]{saenz2011deductive}
S{\'a}enz-P{\'e}rez, F.; Caballero, R.; and Garc{\'\i}a-Ruiz, Y. 2011.
\newblock A deductive database with datalog and sql query languages.
\newblock In \emph{Programming Languages and Systems: 9th Asian Symposium, APLAS 2011, Kenting, Taiwan, December 5-7, 2011. Proceedings 9}, 66--73. Springer.

\bibitem[{Satish et~al.(2010)Satish, Kim, Chhugani, Nguyen, Lee, Kim, and Dubey}]{satish2010fast}
Satish, N.; Kim, C.; Chhugani, J.; Nguyen, A.~D.; Lee, V.~W.; Kim, D.; and Dubey, P. 2010.
\newblock Fast sort on CPUs and GPUs: a case for bandwidth oblivious SIMD sort.
\newblock In \emph{Proceedings of the 2010 ACM SIGMOD International Conference on Management of data}, 351--362.

\bibitem[{Shore(1975)}]{shore1975external}
Shore, J.~E. 1975.
\newblock On the external storage fragmentation produced by first-fit and best-fit allocation strategies.
\newblock \emph{Communications of the ACM}, 18(8): 433--440.

\bibitem[{Shovon et~al.(2023)Shovon, Gilray, Micinski, and Kumar}]{shovon2023towards}
Shovon, A.~R.; Gilray, T.; Micinski, K.; and Kumar, S. 2023.
\newblock Towards iterative relational algebra on the $\{$GPU$\}$.
\newblock In \emph{2023 USENIX Annual Technical Conference (USENIX ATC 23)}, 1009--1016.

\bibitem[{Stonebraker et~al.(2018)Stonebraker, Abadi, Batkin, Chen, Cherniack, Ferreira, Lau, Lin, Madden, O'Neil et~al.}]{stonebraker2018c}
Stonebraker, M.; Abadi, D.~J.; Batkin, A.; Chen, X.; Cherniack, M.; Ferreira, M.; Lau, E.; Lin, A.; Madden, S.; O'Neil, E.; et~al. 2018.
\newblock C-store: a column-oriented DBMS.
\newblock In \emph{Making Databases Work: the Pragmatic Wisdom of Michael Stonebraker}, 491--518.

\bibitem[{Sun et~al.(2023)Sun, Shovon, Gilray, Micinski, and Kumar}]{sun2023gdlog}
Sun, Y.; Shovon, A.~R.; Gilray, T.; Micinski, K.; and Kumar, S. 2023.
\newblock GDlog: A GPU-Accelerated Deductive Engine.
\newblock \emph{arXiv preprint arXiv:2311.02206}.

\bibitem[{Ullman(1983)}]{ullman1983principles}
Ullman, J.~D. 1983.
\newblock \emph{Principles of database systems}.
\newblock Galgotia publications.

\bibitem[{Urbani, Jacobs, and Kr{\"o}tzsch(2016)}]{urbani2016vlog}
Urbani, J.; Jacobs, C.; and Kr{\"o}tzsch, M. 2016.
\newblock Column-oriented datalog materialization for large knowledge graphs.
\newblock In \emph{Proceedings of the AAAI Conference on Artificial Intelligence}, volume~30.

\bibitem[{Urbani et~al.(2010)Urbani, Kotoulas, Maassen, Van~Harmelen, and Bal}]{urbani2010owl}
Urbani, J.; Kotoulas, S.; Maassen, J.; Van~Harmelen, F.; and Bal, H. 2010.
\newblock OWL reasoning with WebPIE: calculating the closure of 100 billion triples.
\newblock In \emph{The Semantic Web: Research and Applications: 7th Extended Semantic Web Conference, ESWC 2010, Heraklion, Crete, Greece, May 30--June 3, 2010, Proceedings, Part I 7}, 213--227. Springer.

\bibitem[{Veldhuizen(2014)}]{veldhuizen2014leapfrog}
Veldhuizen, T.~L. 2014.
\newblock Leapfrog triejoin: A simple, worst-case optimal join algorithm.
\newblock In \emph{Proc. International Conference on Database Theory}.

\bibitem[{Wang, Willsey, and Suciu(2023)}]{wang2023free}
Wang, Y.~R.; Willsey, M.; and Suciu, D. 2023.
\newblock Free join: Unifying worst-case optimal and traditional joins.
\newblock \emph{Proceedings of the ACM on Management of Data}, 1(2): 1--23.

\bibitem[{Weyl et~al.(1975)Weyl, Fries, Wiederhold, and Germano}]{weyl1975modular}
Weyl, S.; Fries, J.; Wiederhold, G.; and Germano, F. 1975.
\newblock A modular self-describing clinical databank system.
\newblock \emph{Computers and Biomedical Research}, 8(3): 279--293.

\bibitem[{Whaley and Lam(2004)}]{whaley2004cloning}
Whaley, J.; and Lam, M.~S. 2004.
\newblock Cloning-based context-sensitive pointer alias analysis using binary decision diagrams.
\newblock In \emph{Proceedings of the ACM SIGPLAN 2004 conference on Programming Language Design and Implementation}, 131--144.

\bibitem[{Zeng et~al.(2023)Zeng, Hui, Shen, Pavlo, McKinney, and Zhang}]{zeng2023empirical}
Zeng, X.; Hui, Y.; Shen, J.; Pavlo, A.; McKinney, W.; and Zhang, H. 2023.
\newblock An Empirical Evaluation of Columnar Storage Formats.
\newblock \emph{Proc. VLDB Endow.}, 17(2): 148--161.

\bibitem[{Zukowski, Nes, and Boncz(2008)}]{zukowski2008dsm}
Zukowski, M.; Nes, N.; and Boncz, P. 2008.
\newblock DSM vs. NSM: CPU performance tradeoffs in block-oriented query processing.
\newblock In \emph{Proceedings of the 4th international workshop on Data management on new hardware}, 47--54.

\end{thebibliography}

\end{document}